\documentclass{PoS}

\title{A search for energy-dependence of the Kes 73/1E 1841-045 morphology in GeV}

\ShortTitle{GeV morphological studies of Kes 73/1E 1841-045}

\author{\speaker{Paul K. H. Yeung}\\
        Institute of Experimental Physics, Department of Physics, University of Hamburg, Luruper Chaussee 149, D-22761 Hamburg, Germany\\
        E-mail: \email{kin.hang.yeung@desy.de}}

\abstract{While the Kes 73/1E 1841-045 system had been confirmed as an extended GeV source, whether its morphology depends on the photon energy or not deserves our further investigation. Adopting data collected by \emph{Fermi} Large Area Telescope (LAT) again, we look into the  extensions of this source in three energy bands individually: 0.3-1 GeV, 1-3 GeV and 3-200 GeV. We find that the 0.3-1 GeV morphology is point-like and is quite different from those in the other two bands, although we cannot robustly reject a unified morphology for the whole LAT band.}

\FullConference{7th Fermi Symposium 2017\\
		15-20 October 2017\\
		Garmisch-Partenkirchen, Germany}

\begin{document}

\section{Introduction}

1E 1841-045 is an anomalous X-ray pulsar (AXP) as well as the central source of the supernova remnant (SNR) Kes 73 \cite{Vasisht1997}. This system shows intense TeV emission at the northern edge of  HESS J1841-055 \cite{Aharonian2008, Bochow2011}. While \cite{Li2017} and \cite{Yeung2017} consistently determined that the GeV counterpart of the Kes 73/1E 1841-045 system is extended with a radius of $\sim$0.32$^\circ$, this is at least four times larger than both radii of the radio shell and the outer X-ray shell (cf. \cite{Kumar2014}\cite{Borkowski2017}). It is interesting to examine whether this morphology is applicable to analyses in the whole LAT band. With regards to this, we break the whole LAT band into three sections for more detailed morphological studies: 0.3-1 GeV, 1-3 GeV and 3-200 GeV. 

\section{Data Analyses and Results}

Aided with \emph{Fermi} Science Tools v10r0p5 package, we reduced and analysed \emph{Fermi} data of a 20$^\circ$$\times$20$^\circ$ region of interest (ROI) centered at RA=$280.28218^\circ$, Dec=$-4.9687344^\circ$ (J2000), which is the centroid of the GeV counterpart of Kes 73/1E 1841-045 \cite{Yeung2017}.  We used the Pass8 ``Clean" class data, accumulated over $\sim$8.9 years. We screened out the ``BACK" data for the sake of better spatial resolution. We further filtered the data by accepting only the good time intervals where the ROI was observed at a zenith angle less than 90$^\circ$ so as to reduce the contamination from the albedo of Earth.

For each energy band, we performed a chain of maximum-likelihood analyses with the improved instrument response function (IRF) ``P8R2$_-$CLEAN$_-$V6". In our source model for background subtraction, we adopted the most updated Fermi/LAT catalog (3FGL \cite{Acero2015a}), which includes the most updated models of Galactic diffuse background (gll$_-$iem$_-$v06.fits) and isotropic background (iso$_-$P8R2$_-$CLEAN$_-$V6$_-$FRONT$_-$v06.txt). We also included the serendipitous sources reported in \cite{Yeung2017}. We  set free the spectral parameters of the  sources within 6$^\circ$ from the ROI center in the analysis. For the   sources beyond 6$^\circ$ from the ROI center, their spectral parameters were fixed at the catalog values. 

\begin{figure}
	\centering
	\includegraphics[width=0.49\linewidth]{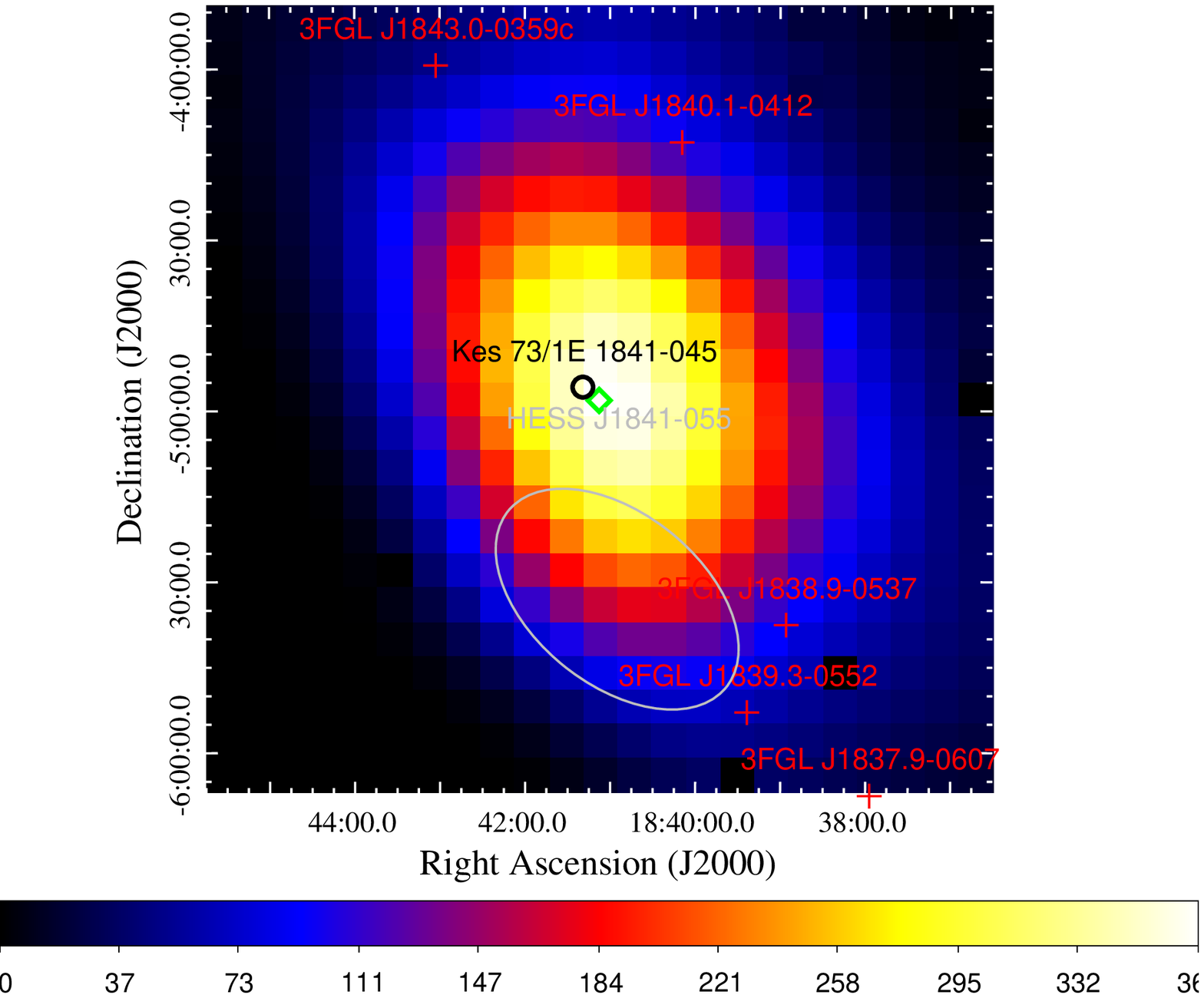}
	\includegraphics[width=0.49\linewidth]{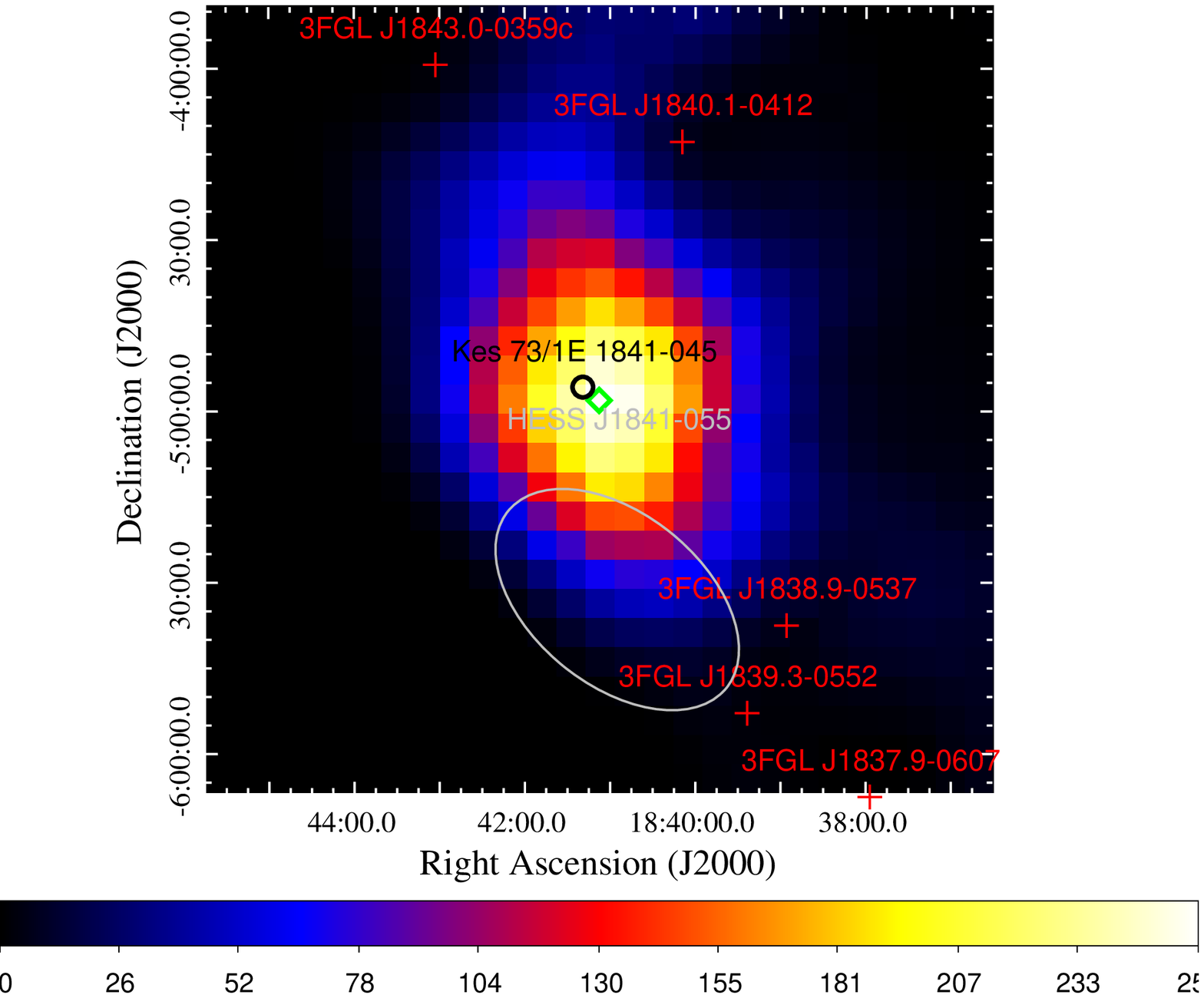}\\
	\includegraphics[width=0.49\linewidth]{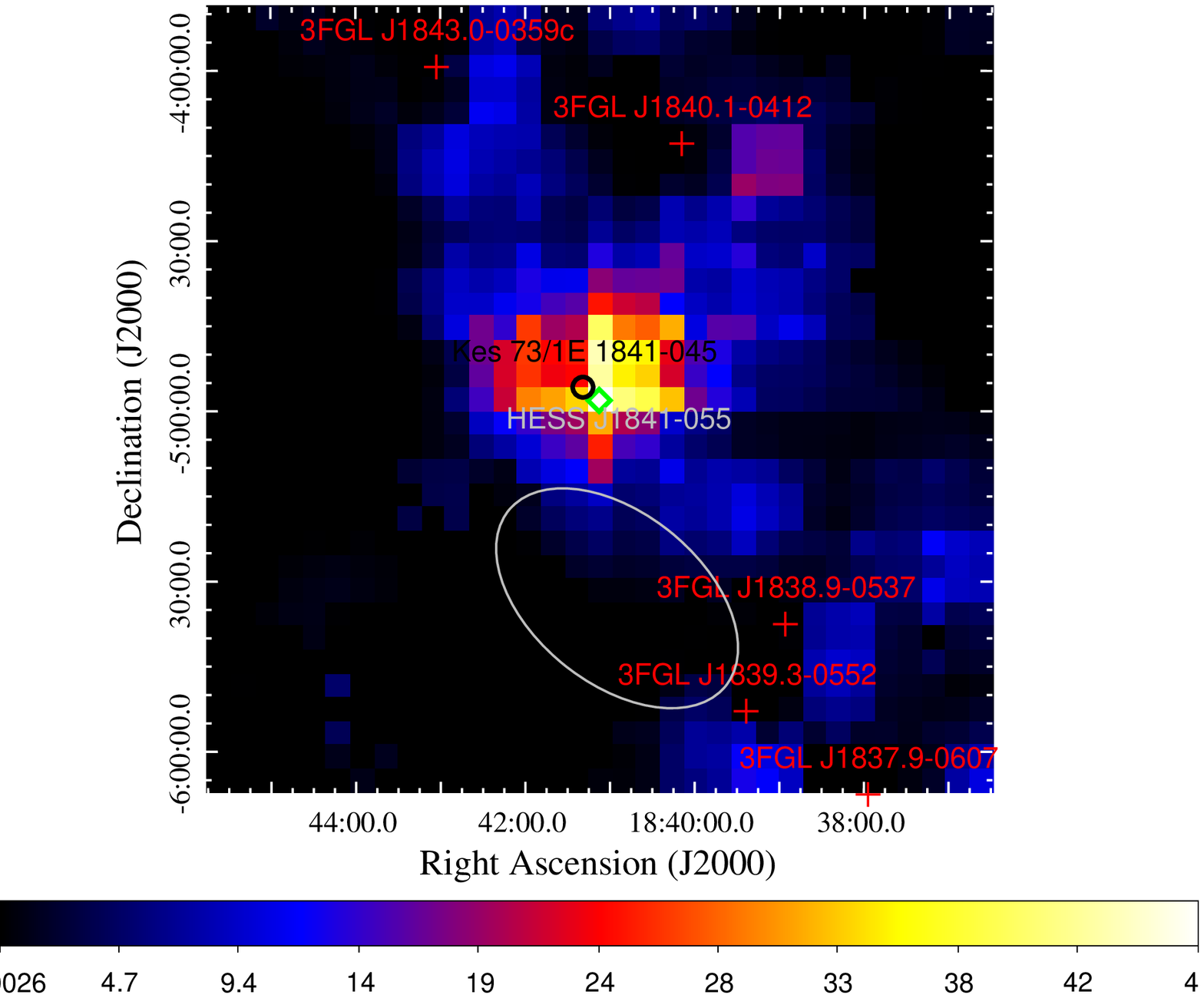}
	\caption{TS maps of the field around 1E 1841-045 at 0.3-1 GeV (top-left), 1-3 GeV (top-right) and 3-200 GeV (bottom) respectively, where all neighboring sources in the model except Kes 73/1E 1841-045 are subtracted. The black thick circle indicates the position and dimension of the Kes 73/1E 1841-045 system in radio, which are taken from \cite{Acero2015b}. The position and dimension of HESS J1841-055,  indicated as a gray ellipse, are taken from \cite{Aharonian2008}.  The positions of nearby 3FGL sources are marked by red crosses. The 1-50 GeV centroid determined in \cite{Yeung2017} is indicated as a green diamond.}
	\label{1E_tsmap}
\end{figure}

The test-statistic (TS) maps, at 0.3-1 GeV, 1-3 GeV and 3-200 GeV respectively, are shown in Figure~\ref{1E_tsmap}, where all sources in the model except Kes 73/1E 1841-045 are subtracted. As we can see, the $\gamma$-ray features in these three bands share a common centroid, which is also the same as the 1-50 GeV centroid reported in \cite{Yeung2017}. Hence, we continued to take this centroid position for investigating the extensions of the Kes 73/1E 1841-045 associated source in both bands.

In each band, we  performed a likelihood ratio test to quantify the radius and significance of extension, following the scheme adopted by  \cite{Yeung2016} and \cite{Yeung2017}. We assigned this source a simple power law, and we attempted  uniform-disk morphologies of different radii as well as a point-source model on it. The results are tabulated in Table~\ref{Ext}.
The 0.3-1 GeV morphology can be described by either a point model or an extended model of $<0.35^\circ$ radius, within the tolerance of the $2\sigma$ uncertainty.
On the other hand, the 1-3 GeV morphology is extended with a $0.369^\circ\pm0.040^\circ$ radius at a $>6\sigma$ level, and the 3-200 GeV morphology is extended with a $0.383^\circ\pm0.046^\circ$ radius at a $>4.5\sigma$ level.

\begin{table}
	\centering
	\caption{Morphological properties of the Kes 73/1E 1841-045 associated source in different energy bands.}
	\label{Ext}
	\begin{tabular}{ccc}
		\hline\hline
		Energy Band & Radius of Extension & $2ln(L_{ext}/L_{pt})$ \\
		(GeV)       & (deg)              &               \\ \hline
		0.3-1       & 0.167$\pm$0.091 & 1.30          \\
		1-3         & 0.369$\pm$0.040 & 37.13         \\
		3-200       & 0.383$\pm$0.046 & 21.89       \\ \hline
	\end{tabular}
\end{table}

\section{Discussion and Conclusion}

While it is unascertainable whether the actual morphology in 0.3-1 GeV is a point or an extended clump, its allowable extension is smaller than each of the 1-3 GeV and 3-200 GeV morphologies by $>2\sigma$. Since the 95\% upper limit of the radius of extension in 0.3-1 GeV is $0.35^\circ$, the spectral energy distribution in \cite{Yeung2017} with a unified radius of extension of $0.32^\circ$ could still be a rough approximation for the 0.3-1 GeV band. Nevertheless, we encourage ones to look into the 0.3-1 GeV properties of this source by assigning it a point morphology or a less extended morphology, so that we can distinguish between Kes 73 and 1E 1841-045 contributions more accurately. The systematic uncertainties induced by the error of extension should also be taken into account. On the other hand, the 1-3 GeV and 3-200 GeV morphologies are in a good agreement, within the tolerance of statistical uncertainties.

We recall that the position and dimension of HESS J1841-055 has not been refined yet. Thus, the interplay between the morphologies of the Kes 73/1E 1841-045 associated source and HESS J1841-055 still remains a crucial issue in this work as well as in \cite{Yeung2017}. 

Noticeably, \cite{Ackermann2017} replaced HESS J1841-055 with two extended sources FGES J1839.4-0554 and FGES J1841.4-0514 in their $>$10 GeV analyses, where the latter is spatially consistent with our targeted AXP/SNR pair. This is undoubtedly an alternative way to goodly describe the entire $\gamma$-ray complex. Whereas, FGES J1841.4-0514 contains other high-energy components as well. Since we intend to model the AXP/SNR pair as a single source independent from other high-energy sources, their revised morphological model for this field is not applicable to our studies.

\end{document}